\documentclass[journal,twocolumn]{IEEEtran}
\usepackage{graphicx, subfigure}
\usepackage{caption}
\usepackage{makecell}
\usepackage{soul}
\usepackage{listings}
\usepackage{multirow}
\usepackage{pifont}
\usepackage{xspace}
\usepackage{epigraph} 
\usepackage[flushleft]{threeparttable}
\usepackage{url}
\usepackage{algorithm2e}
\UseRawInputEncoding
\usepackage[colorinlistoftodos]{todonotes}
\usepackage{listings}
\usepackage{graphicx,wrapfig}
\usepackage{comment}
\usepackage{fontawesome}
\usepackage{booktabs}
\definecolor{celestialblue}{rgb}{0.29, 0.59, 0.82}
\definecolor{awesome}{rgb}{0.0, 0.2, 0.6}
\definecolor{coolblack}{rgb}{0.0, 0.18, 0.39}
\definecolor{maroon}{cmyk}{0, 0.87, 0.68, 0.32}
\definecolor{halfgray}{gray}{0.55}
\definecolor{ipython_frame}{RGB}{207, 207, 207}
\definecolor{ipython_bg}{RGB}{247, 247, 247}
\definecolor{ipython_red}{RGB}{186, 33, 33}
\definecolor{ipython_green}{RGB}{0, 128, 0}
\definecolor{ipython_cyan}{RGB}{64, 128, 128}
\definecolor{ipython_purple}{RGB}{170, 34, 255}
\usepackage[strict]{changepage}
  \definecolor{ABlue}{HTML}{127bca}
 \definecolor{LHScolor}{HTML}{555555}
\usepackage{framed}

\definecolor{formalshade}{rgb}{1.0,1.0,1.0}
\definecolor{side}{rgb}{0.0,0.2,0.6}
\usepackage[skins,breakable]{tcolorbox}
\definecolor{Large}{HTML}{696969}
\definecolor{Negligible}{HTML}{D3D3D3}
\definecolor{Medium}{HTML}{808080}
\definecolor{Small}{HTML}{A9A9A9}
\definecolor{backcolour}{rgb}{0.95,0.95,0.92}

\definecolor{chestnut}{rgb}{0.8, 0.36, 0.36}

\definecolor{chestnut}{rgb}{0.8, 0.36, 0.36}

\usepackage[skins,breakable]{tcolorbox}

\definecolor{Large}{HTML}{696969}
\definecolor{Negligible}{HTML}{D3D3D3}
\definecolor{Medium}{HTML}{808080}
\definecolor{Small}{HTML}{A9A9A9}

\begin{document}

\newtheorem{theorem}{Definition}[section]
	\newcommand{\eg}{e.g.,}
	\newcommand{\ie}{i.e.,}	
	\renewcommand{\lstlistingname}{Listing}

\newcommand{\boxedtext}[1]{\fbox{\scriptsize\bfseries\textsf{#1}}}
\newcommand{\nota}[2]{
	\boxedtext{#1}
		{\small$\blacktriangleright$\emph{\textsl{#2}}$\blacktriangleleft$}
}

\newcommand\review[3]{\textcolor{red}{\sout{#1}} {\textcolor{blue}{#2}}{\todo{#3}}}

\definecolor{beaublue}{rgb}{0.74, 0.83, 0.9} 
\definecolor{redbeau}{HTML}{cc5b5b}  
\definecolor{orabeau}{HTML}{ff9f79}

\title{The Impact of Sanctions \\ on GitHub Developers and Activities}

\author{
\IEEEauthorblockN{Youmei Fan\IEEEauthorrefmark{1}, 
Ani Hovhannisyan\IEEEauthorrefmark{1}, 
Hideaki Hata\IEEEauthorrefmark{2}
Christoph Treude\IEEEauthorrefmark{3}}
 and
 Raula Gaikovina Kula\IEEEauthorrefmark{1}
\\
\IEEEauthorblockA{
\IEEEauthorrefmark{1}NAIST, Japan,
\IEEEauthorrefmark{2}Shinshu University, Japan,
\IEEEauthorrefmark{3}Singapore Management University, Singapore
\\
fan.youmei.fs2@is.naist.jp, hovhannisyan.ani.hb7@is.naist.jp, hata@shinshu-u.ac.jp, ctreude@smu.edu.sg, raula-k@is.naist.jp}
}

\maketitle

\begin{abstract}
The GitHub platform has fueled the creation of truly global software, enabling contributions from developers across various geographical regions of the world. 
As software becomes more entwined with global politics and social regulations, it becomes similarly subject to government sanctions. In 2019, GitHub restricted access to certain services for users in specific locations but rolled back these restrictions for some communities (e.g., the Iranian community) in 2021. 
We conducted a large-scale empirical study, collecting approximately 156 thousand user profiles and their 41 million activity points from 2008 to 2022, to understand the response of developers.
Our results indicate that many of these targeted developers were able to navigate through the sanctions.
Furthermore, once these sanctions were lifted, these developers opted to return to GitHub instead of withdrawing their contributions to the platform. 
The study indicates that platforms like GitHub play key roles in sustaining global contributions to Open Source Software. 
\end{abstract}

\begin{keywords}
Government Sanctions, Open Source
\end{keywords}
\vspace{-1em}

\section{Introduction}
Much of GitHub's success can be attributed to the digital age, where technology transcends geographical borders. Due to its open availability, GitHub is home to many popular OSS ecosystems, including NPM, PyPI, and Maven, to name a few~\cite{SofEco2023}.
GitHub has acknowledged the difficulty of complying with international policies while maintaining a global community, particularly for developers impacted by societal issues related to their geographical location. An example is the rise of protestware~\cite{kula2022war}, which targets specific communities and software.

Government sanctions, often inevitable, are a common international response to issues such as armed conflicts, human rights abuses, and terrorism concerns. For example, the United States (US) imposes broad country-level prohibitions~\cite{treasuryHome} and maintains a list of Specially Designated Nationals and Blocked Persons, including individuals and entities facing comprehensive sanctions~\cite{treasurygov}. Previous research attests to the significant impact of these sanctions~\cite{bezuidenhout2019economic}.
According to GitHub:

\epigraph{
\color{coolblack}{\textit{``\textit{Complying with these sanctions isn't a choice based on what we (GitHub) think about a particular country or the developers in it. ...We implemented access restrictions for developers we understand to be located or resident in sanctioned countries, and not based on nationality or heritage.}'' }}}{--\textit{GitHub} \faGithub
}
Originally, sanctions were designed to regulate the trade of traditional goods and services, especially financial products. 
However, the rise of digital services has extended the reach of these regulations into the software engineering domain. 
In 2021, GitHub announced that it had lifted the sanctions imposed on Iran~\cite{advancingdev}:

\epigraph{
\color{coolblack}{\textit{``\textit{Today we are announcing a breakthrough: we have secured a license from the US government to offer GitHub to developers in Iran. This includes all services for individuals and organizations, private and public, free and paid.}''}}}{--\textit{GitHub} \faGithub}

\begin{figure}[t]
    \centering
    \includegraphics[width=1\linewidth]{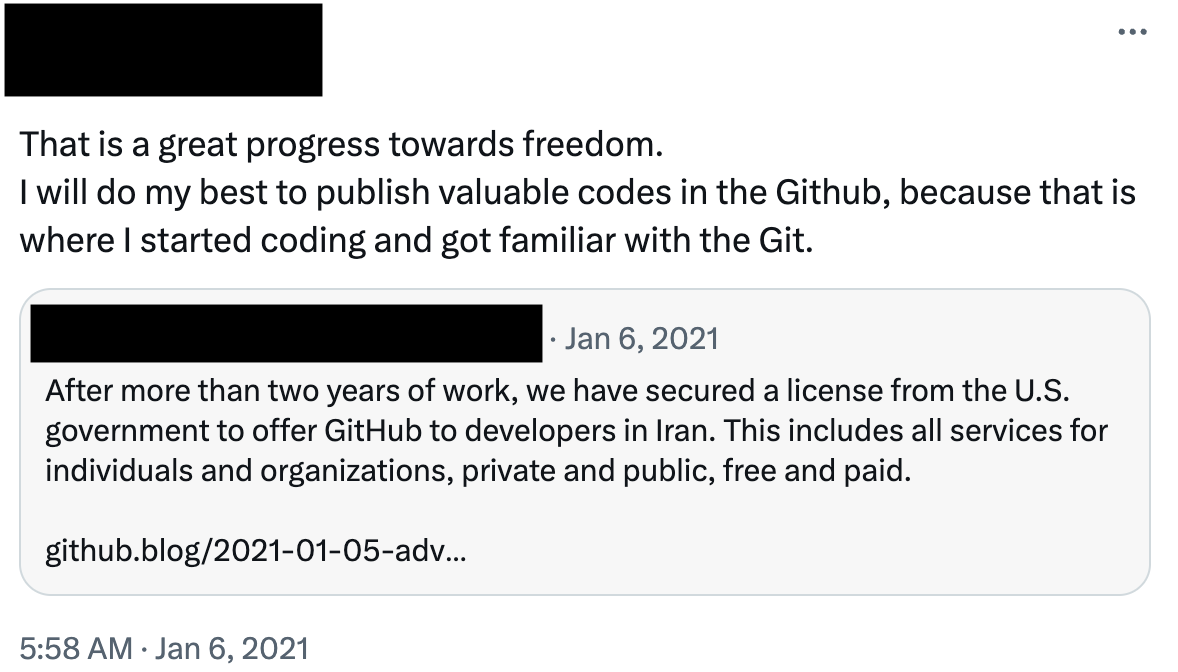}
    \caption{A GitHub developer's reaction to the announcement of the Iranian sanctions been lifted.}
    \label{fig:overview}
\end{figure}

\begin{table*}[t]
    \centering
    \caption{Number of Users and Activities Across Countries and Regions}
    \begin{tabular}{lrrrrrrrr|r}
    \hline
         & {Iranian} & {Crimean} & {Cuban} & {Syrian} & {Russian} & {Greek} & {Kenyan} & {Hong Kongese} & {Total} \\ \hline
        {\# Registered Users} & 26,333 & 200 & 1,711 & 1,456 & 77,896 & 17,047 & 17,409 & 14,441 & 156,493 \\
        {\# GitHub Activities} & 3,148,280 & 34,684 & 391,179 & 98,864 & 17,966,530 & 6,200,152 & 7,755,674 & 5,662,327 & 41,257,690 \\ \bottomrule
    \end{tabular}
    \label{tab:DataRQ1}
\end{table*}

Fig. 1 illustrates a developer's reaction to the lifting of the Iranian sanctions\footnote{ 
For privacy reasons, the developer's identity is kept confidential.}. 
In response to the announcement of the sanctions being lifted, the developer expresses gratitude and explicitly states that such actions represent \textit{progress towards freedom} as well as the intention of contributing code to GitHub.

This event offers a unique opportunity to longitudinally examine the impact on developers affected by the sanctions on the platform.
At the same time, GitHub has been expanding its services to other regions of the world, such as Asia, Africa, and Europe, fostering greater inclusively.
GitHub also stated:

\epigraph{
\color{coolblack}{\textit{``\textit{Outside of these regions, we also saw a continued rise in Iranians logging into GitHub (a 21\% year-over-year increase, to be precise) following our work from 2019 through 2021 to secure a license from the US government to offer service to Iranian developers. And we applaud the bravery of the Iranian people, especially women, who are putting themselves at risk in pursuit of freedom.}'' }}}{--\textit{GitHub} \faGithub}

To better understand the implications of sanctions on GitHub developers and activities, we performed an empirical investigation from two perspectives. 
For the first perspective, we concentrate on general GitHub activities.
We have collected data on 156,493 user profiles from eight countries\footnote{Throughout this paper, the term ``countries'' is used to refer to both countries and regions.}: Iran, Crimea, Cuba, Syria, Russia, Greece, Kenya, and Hong Kong. 
We use some of these countries as baselines to compare the affected countries with those that were not impacted.
Our findings indicate that sanctioned communities were able to find ways to remain active on the platform. 
For the second perspective, we concentrate on the case of developers in Iran and whether they returned after the sanctions were lifted. 
We investigated how many of the developers stopped their GitHub activities and to what extent the activities returned to normal after the sanctions were lifted.

\section{Data Preparation}

\subsection{\textbf{Target Country Selection}}
To investigate the impact of sanctions on GitHub developers and activities, we collected data from several countries and regions. In addition to Iran, we selected countries and regions that were sanctioned at the same time. According to GitHub, the sanctioned list includes Crimea, the separatist areas of Donetsk and Luhansk, Cuba, Iran, North Korea, and Syria. Due to time constraints and the practicality of the analysis, we sampled four of the six sanctioned areas.
As baselines, we selected Russia, Greece, Kenya, and Hong Kong. Russia was chosen because of its absence of GitHub sanctions, despite having contentious relationships with the US. Other communities, such as Kenya, Greece, and Hong Kong, were found to have growth trajectories similar to those of Iran~\cite{octoverse2019,octoverse2020}. Additionally, due to the overlap of the sanction period with the COVID-19 pandemic, establishing a baseline helps us accurately assess the impact of the sanctions without confusing it with the effects of the pandemic.

\subsection{\textbf{Data Sources}}

We used the GitHub API and the Graph Query to collect data for developers from the target countries. To search and identify developers, we used the {location} and {homepage} attributes in the {user} query in the API~\cite{freeproteam}. To collect their GitHub activities, we acquired information regarding users' annual contributions by using the GitHub Query Language:
\begin{quote}
\texttt{query:"location:xxx type:user created:the\_day", type: USER, first: 100}
\end{quote}

Note that, while collecting user contributions, some users had already canceled their accounts, or their account type was set to private. In these cases, we omitted them from the subsequent analysis. We limited our search to users registered between 1 April 2008 and 1 May 2023. Since we were only interested in active users, we removed all users who did not have any contributions in 2022. To facilitate this large-scale study, we only collected the total annual count of contributions per year for each user.

Table \ref{tab:DataRQ1} shows a summary of the data collected for each country. Russia has the highest number of total users, while Crimea has the fewest. In total, we collected 156,493 users. In terms of activities, Russia has more activities than the rest. Iran has the highest activity among sanctioned communities, while users from Greece, Kenya, and Hong Kong accounted for approximately 5.5 to 7.7 million activities.

\begin{figure*}[t]
    \centering
    \includegraphics[width=.6\linewidth]{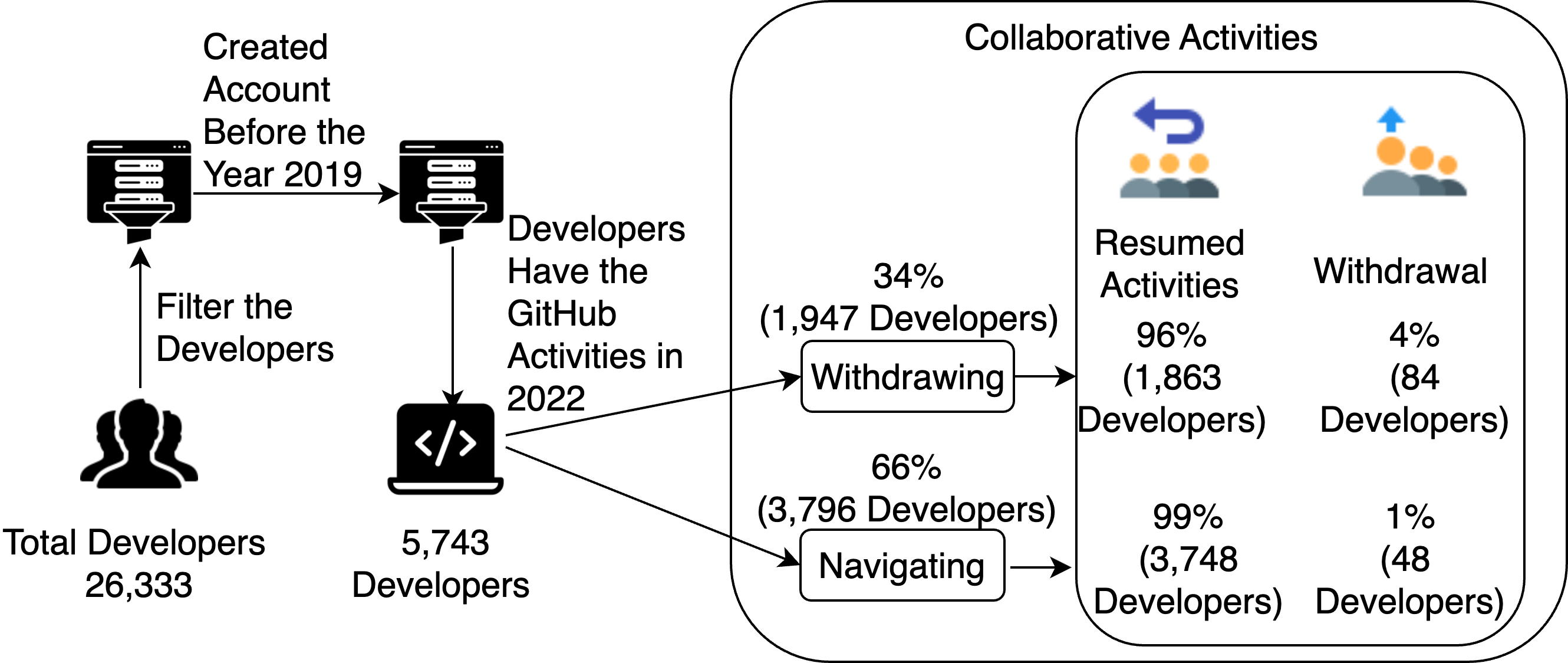}
    \caption{Flow diagram depicting sanctioned developers who returned to GitHub after the sanctions were lifted}
    \label{fig:overview}
\end{figure*}

\section{Tracking Activities of Sanctioned Developers}

To measure the change in activities, we used the relative change metric (or percent error), i.e., the relative change expressed as a percentage~\cite{boos2013essential}, in activities for each year between 2012 and 2022. It is used to compare two quantities while considering the `sizes' of the things being compared, expressing the terms of percentage change. We utilized a time series to observe these changes over time.
To evaluate a change in 
users and their activities, we examined the trend up to 2019 and then those activities after 2020. This aligns with the sanctions that took place from July 2019 to January 2021; therefore, our analysis focused on the relative change percentage from 2019 to 2020.

\begin{table}[]
    \centering
    \caption{Relative change percentage (mean values) for each community. 
    }
    \label{tab:RQ1a}
    \begin{tabular}{lrrrrrrrr}
    \hline
        Year & 2018 & 2019 & 2020 & 2021 & 2022 \\ \hline
        Iran & 40.7 & \fcolorbox{black}{green}{152.3} & \fcolorbox{black}{green}{260.4} & \fcolorbox{black}{green}{689.9} & \fcolorbox{black}{green}{857.2} \\
        Crimea & 75.8 & \fcolorbox{black}{pink}{3.1} & \fcolorbox{black}{green}{625.0} & \fcolorbox{black}{pink}{21.5} & \fcolorbox{black}{green}{1, 364.4} \\
        Cuba & 150.0 & \fcolorbox{black}{pink}{141.0} & \fcolorbox{black}{green}{638.5} & \fcolorbox{black}{pink}{568.9} & \fcolorbox{black}{green}{778.6} \\         
        Syria & 45.2 & \fcolorbox{black}{pink}{35.5} & \fcolorbox{black}{green}{67.4} & \fcolorbox{black}{green}{338.7} & \fcolorbox{black}{green}{723.5} \\
        Russia & 148.0 & \fcolorbox{black}{green}{266.7} & \fcolorbox{black}{green}{438.9} & \fcolorbox{black}{green}{620.4} &\fcolorbox{black}{green}{700.2} \\ 
        Greece & 130.7 & \fcolorbox{black}{green}{189.6} & \fcolorbox{black}{green}{348.7} & \fcolorbox{black}{green}{615.2} & \fcolorbox{black}{pink}{508.1} \\ 
        Kenya & 190.3 & \fcolorbox{black}{green}{341.1} & \fcolorbox{black}{green}{509.4} & \fcolorbox{black}{green}{708.8} & \fcolorbox{black}{green}{2, 267.4} \\
        Hong Kong & 162.9 & \fcolorbox{black}{green}{216.0} & \fcolorbox{black}{green}{376.7} & \fcolorbox{black}{green}{571.7} & \fcolorbox{black}{green}{624.0} \\ \hline
    \end{tabular}
\end{table}

Table \ref{tab:RQ1a} shows the rate of GitHub activities for each country from 2018 to 2022.
We have annotated the table to indicate a rise in green color or a drop in pink color.
All communities experienced an increase in GitHub activities in 2020.
Especially for developers in Iran, we see a growth median change percent rise from 152.3 to 260.4 in 2019.
As illustrated in Table \ref{tab:RQ1a}, the three communities that decreased their activities once the sanctions started were Crimea, Cuba, and Syria, experiencing a drop in activities from a mean of 75.8 to 3.1, 150 to 141 and 45.2 to 35.5
respectively. 
This trend is expected, as sanctions against these communities have been applied. 

\begin{tcolorbox}[colback=gray!5,colframe=awesome,title= Summary]

Developers from the sanctioned communities have resumed their activities as usual in, finding a way to navigate the sanctions. Crimea, Cuba, and Syria declined in activity at the start of the sanctions period. Additionally, Iranian users showed continuous growth from 2019 to 2020, emphasizing GitHub's crucial role in fostering open-source collaboration under restrictions.
\end{tcolorbox}

\section{The Case when Sanctions were Lifted }

To understand how collaborative activities are impacted when sanctions are lifted, we examine developers in Iran, since they recently had their restrictions removed.
Using the same dataset, we identify and collect developers who have set the location in their profile as being from Iran.
To identify only developers who would be impacted by the sanction, we omit developers who registered after 2019. 
After filtering, we identified {5,743} users. 
Based on the GitHub API and our own web analysis, we were able to distinguish contributions into two groups:
\begin{itemize}
    \item \textbf{Collaborative Activities} - These contributions are defined as potential collaborations with other users on GitHub.\footnote{https://docs.github.com/en/graphql/reference/objects\#contributionscollection} Contributions include creating a repository, making a pull request, reviewing a pull request, committing to the repositories, and opening an issue.
    \item \textbf{GitHub Activities} - Activities include joining an organization, starting a discussion in their own repositories, etc. These activities indicate personal activities on users' own repositories. 
\end{itemize}

Figure \ref{fig:overview} shows the different pathways of affected developers and shows how many of them returned to the platform once the sanctions were lifted. We verify users' GitHub activities by checking whether they have at least one GitHub activity during and after the sanctions had been lifted.  
We use the term navigation to identify those developers who found a way to make contributions despite the sanctions.
Although it is not clear how developers were able to bypass the restrictions, 3,748 developers found a way to make a contribution during the sanctions and continued to do so after the sanctions were lifted.

In this study, we find that 34\% of users were impacted and discontinued their contributions once the sanctions came into effect.
However, as shown in Figure \ref{fig:overview}, many returned after the sanctions were lifted. 
Only a small proportion of developers permanently withdraw from the platform. 
In fact, of the {1,947} developers who ceased activities, only {4}\% have not returned to GitHub since the sanctions were enforced. 

\begin{tcolorbox}[colback=gray!5,colframe=awesome,title= Summary]
Although developers find ways to navigate the sanctions, the results show that the lifting of sanctions by GitHub did correlate with the resumed activities of the sanctioned developers. Of the 5,743 developers, 34\% temporarily withdrew from making contributions; however, almost 96\% of the developers who had temporarily withdrawn returned once the sanctions were removed. 

\end{tcolorbox}

\section{Limitations}

In this section, we discuss the limitations and future outlook and implications of the results found in this study. 
A comprehensive empirical investigation is required to thoroughly examine the effects of sanctions, which would involve conducting a survey among developers.
However, given the delicate nature of the subject, it seems improbable that such a study will be possible.
We anticipate that future research will build on our findings and extend the research to additional regions.
We recognize that our approach to systematically identifying GitHub users associated with specific regions is vulnerable to inaccuracies caused by users incorrectly listing their locations.
To address this issue, we collected a substantial volume of data, mitigating the impact of such inaccuracies through statistical means.

\section{Conclusion}
 The lifting of sanctions against developers in Iran on GitHub provided a unique opportunity to study trends on how developers from sanctioned regions of the world cope with sanctions.
Our results are mixed. On one hand, sanctioned developers continued logging into GitHub and conducting their regular activities; on the other hand, there was evidence that the lifting of sanctions correlated with resumed activities among these developers, who returned to contributions once sanctions were removed.
Future detailed studies may be needed to confirm whether the lifted sanctions were responsible for the increase in developers, as this trend was seen in all communities.

\section{Acknowledgement}
This work was supported by JST SICORP Renewmap Grant Number JPMJSC2206.
The work is also supported by  the Japanese Society for the Promotion of Science (JSPS) KAKENHI grants (JP23K16864, JP20H05706).

\bibliographystyle{IEEEtran}
\bibliography{filtered_ref}
\vspace{1\baselineskip}

{\setlength\intextsep{0pt}
\begin{wrapfigure}{l}{25mm} 
    \includegraphics[width=1in,height=1.25in,clip,keepaspectratio]{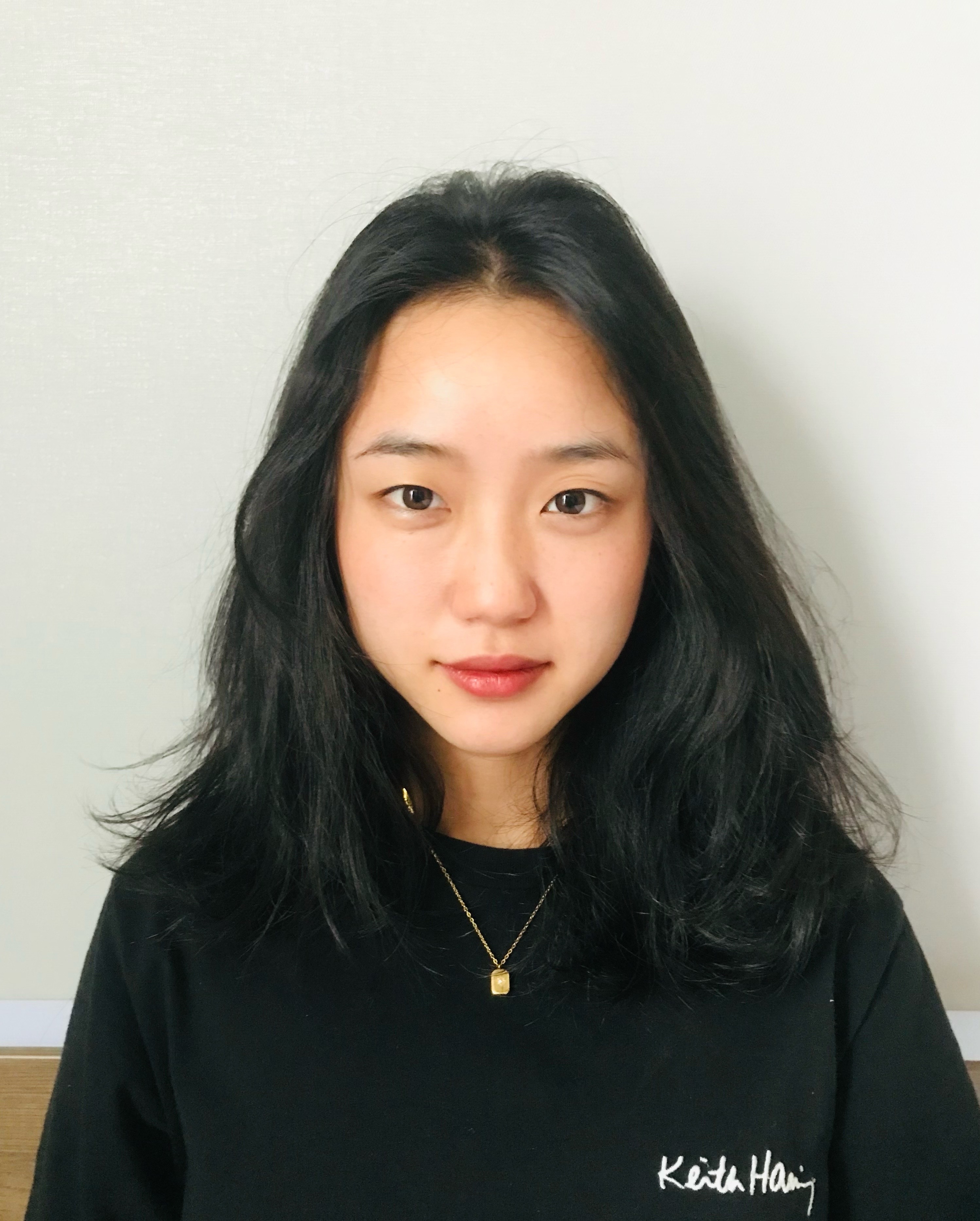}
\end{wrapfigure}\par
\noindent\textbf{Youmei Fan}\\ is a Ph.D. student in Software Engineering Laboratory at the Graduate School of Science and Technology, Nara Institute of Science and Technology (NAIST). She completed a master's course in Software Engineering Laboratory at the Graduate School of Science and Technology, Nara Institute of Science and Technology (NAIST), Japan. During her master's degree, her main research interests related to the human aspect of software engineering, open-source software sustainability, and data mining.
Find her at \url{https://www.linkedin.com/in/youmei-fan-513a161a6/}.
\par}

\vspace{1\baselineskip}

{\setlength\intextsep{0pt}
\begin{wrapfigure}{l}{25mm} 
    \includegraphics[width=1in,height=1.25in,clip,keepaspectratio]{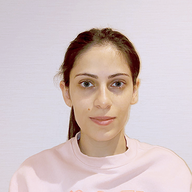
}
\end{wrapfigure}\par
\noindent\textbf{Ani Hovhannisyan}\\ is a Ph.D. student in the Software Engineering Laboratory at the Graduate School of Science and Technology, Nara Institute of Science and Technology (NAIST), Japan. She is interested in mining repositories. \par}

\vspace{1\baselineskip}

{\setlength\intextsep{0pt}
\begin{wrapfigure}{l}{25mm} 
    \includegraphics[width=1in,height=1.25in,clip,keepaspectratio]{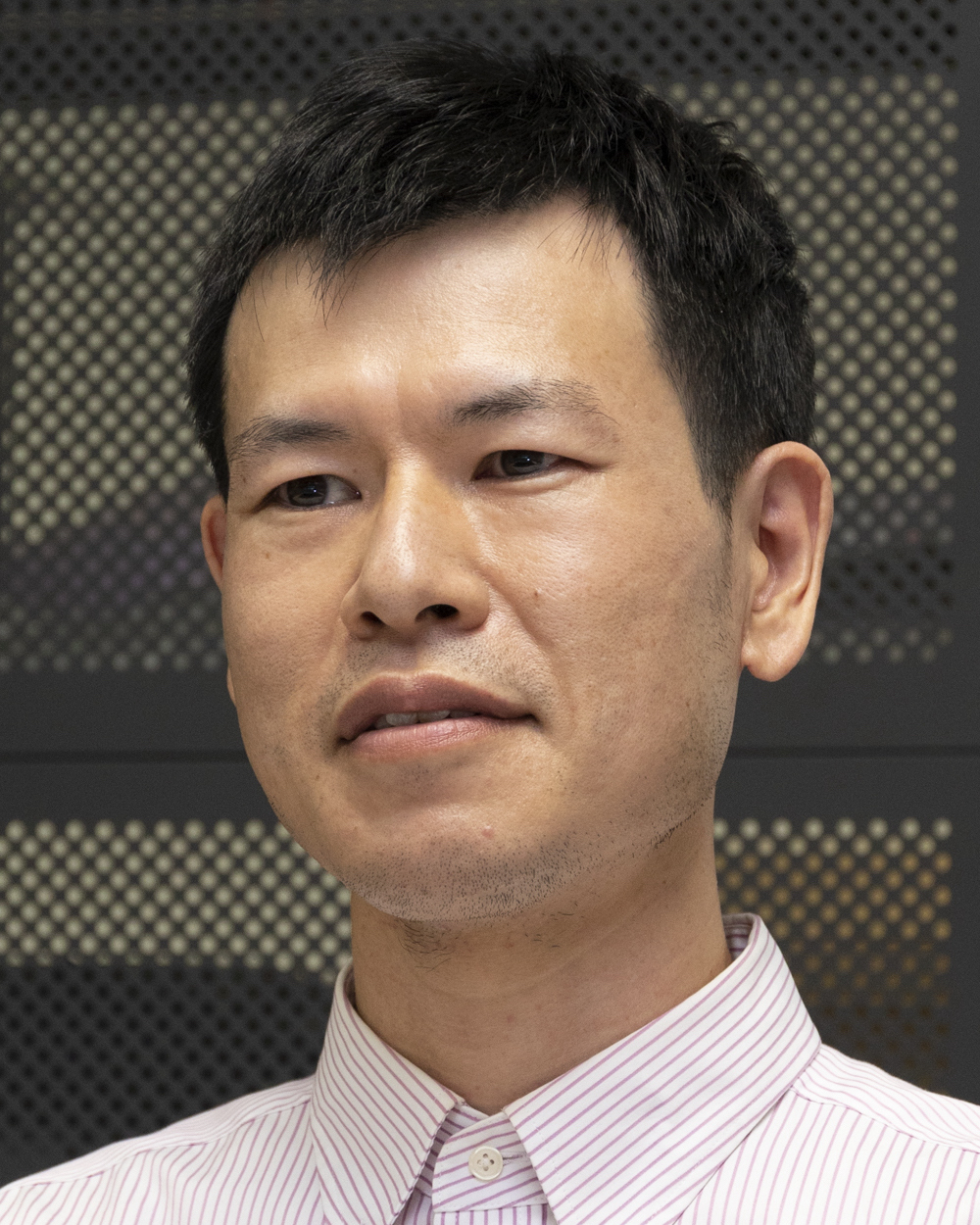}
\end{wrapfigure}\par
\noindent\textbf{Hideaki Hata}\\ is an Associate Professor at Shinshu University. He received his Ph.D. in information science from Osaka University. His research interests include software ecosystems, human capital in software engineering, and software economics. More about Hideaki and his work is available online at \url{https://hideakihata.github.io/}. \par}

\vspace{1\baselineskip}

{\setlength\intextsep{0pt}
\begin{wrapfigure}{l}{25mm} 
    \includegraphics[width=1in,height=1.25in,clip,keepaspectratio]{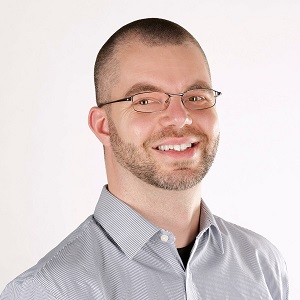}
\end{wrapfigure}\par
\noindent\textbf{Christoph Treude}\\ is an Associate Professor of Computer Science at Singapore Management University. He has authored over 150 scientific articles with more than 250 co-authors. His work has received recognition, including an ARC Discovery Early Career Research Award (2018-2020) and funding from industry leaders such as Google, Facebook, and DST. Treude has received four best paper awards, including two ACM SIGSOFT Distinguished Paper Awards. Currently, Treude serves on the Editorial Boards of the IEEE Transactions on Software Engineering, the Springer Journal on Empirical Software Engineering, and the Wiley Journal of Software: Evolution and Process. He also holds the role of Open Science Editor for the Elsevier Journal of Systems and Software. He has chaired conferences such as ICSME 2020, ICPC 2023, and TechDebt 2023 and regularly participates in software engineering conference program committees. \par}

\vspace{1\baselineskip}

{\setlength\intextsep{0pt}
\begin{wrapfigure}{l}{25mm} 
    \includegraphics[width=1in,height=1.25in,clip,keepaspectratio]{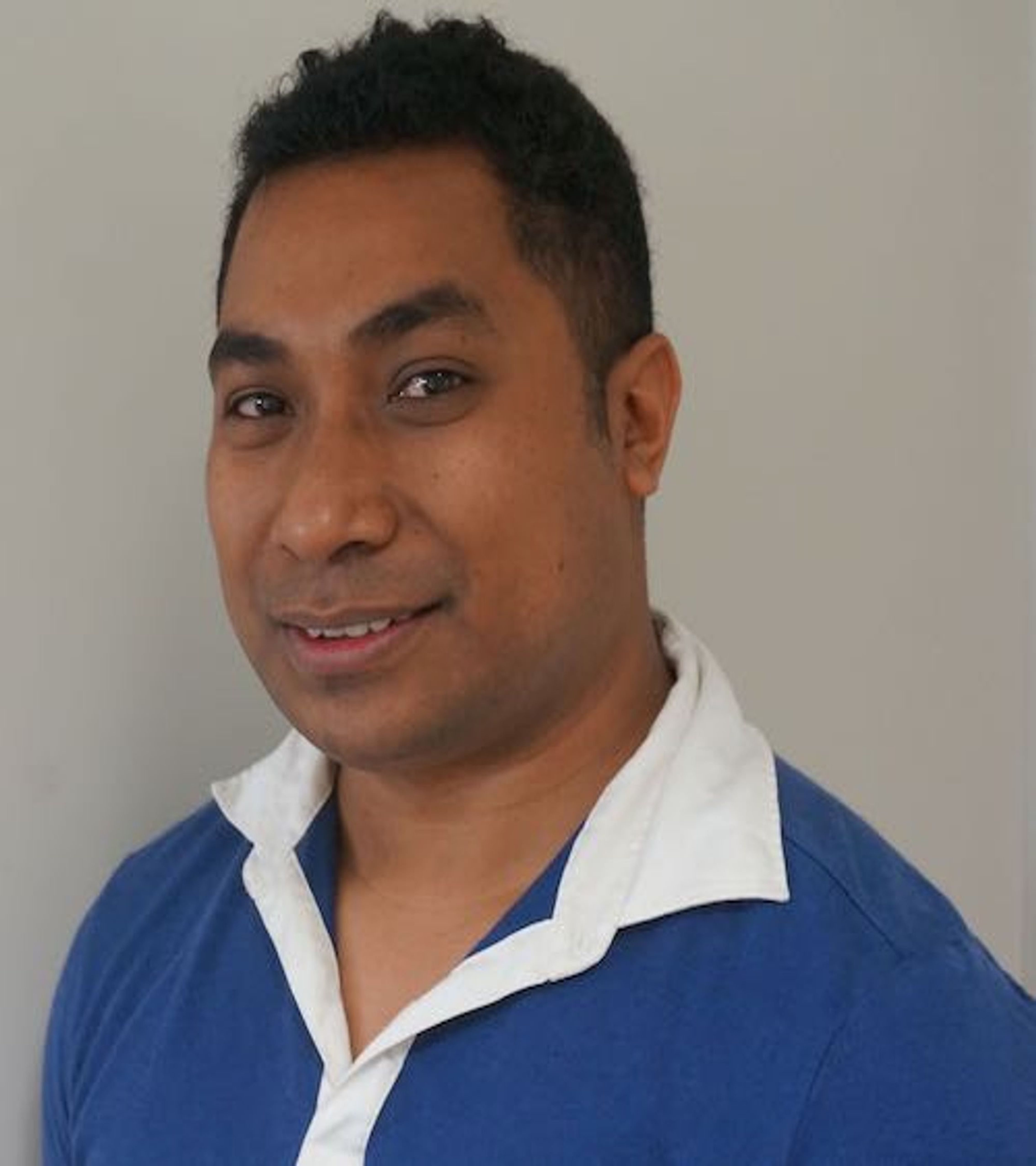}
\end{wrapfigure}\par
\noindent\textbf{Raula Gaikovina Kula}\\ is an Associate Professor at the Nara Institute of Science
and Technology (NAIST), Japan. He received his Ph.D. degree from NAIST in 2013. He is active in the
Software Engineering community, serving the community as a PC member for
premium SE venues, some as organising committee, and reviewer for journals.
His current research interests include library dependencies and security in the
software ecosystem, program analysis such as code clones, and human aspects such as code reviews and coding proficiency. Find him at \url{https://raux.github.io/}
and @augaiko on Twitter. Contact him at raula-k@is.naist.jp. \par}

\end{document}